\documentclass[lettersize,journal]{IEEEtran}

\usepackage[T1]{fontenc}
\usepackage{mathrsfs}
\usepackage{bm}
\usepackage{amsmath}
\usepackage{amsthm}
\usepackage{amssymb}
\usepackage{float}
\usepackage{graphicx}
\usepackage{verbatim}

\usepackage{bbm}
\graphicspath{{Images/}}
\PassOptionsToPackage{normalem}{ulem}

\usepackage{cite}
\usepackage{graphicx,psfrag,cite,subfigure}
\usepackage[table]{xcolor}

\newcommand\scalemath[2]{\scalebox{#1}{\mbox{\ensuremath{\displaystyle #2}}}}
\DeclareMathOperator{\esssup}{ess\: sup}
\setkeys{Gin}{width=1.0\columnwidth}

\begin{document}
\title{Quickest Detection of False Data Injection Attack in Distributed Process Tracking

\thanks{Saqib Abbas Baba is with the Department of Electrical Engineering, Indian Institute of Technology (IIT), Delhi. Email: saqib.abbas@ee.iitd.ac.in}
\thanks{Arpan Chattopadhyay is with the Department of Electrical Engineering and the Bharti School of Telecom Technology and Management, Indian Institute of Technology (IIT), Delhi. Email: arpanc@ee.iitd.ac.in}
\thanks{A. C. acknowledges support via the professional development fund and professional development  allowance from IIT Delhi, grant no. GP/2021/ISSC/022 from I-Hub Foundation for Cobotics and grant no. CRG/2022/003707 from Science and Engineering Research Board (SERB), India.}
}

\author{
Saqib Abbas Baba, Arpan Chattopadhyay}

\maketitle
\ifdefined\SINGLECOLUMN
	\setkeys{Gin}{width=0.5\columnwidth}
	\newcommand{\figfontsize}{\footnotesize} 
\else
	\setkeys{Gin}{width=1.0\columnwidth}
	\newcommand{\figfontsize}{\normalsize} 
\fi

\begin{abstract} 
This paper addresses the problem of detecting false data injection (FDI) attacks in a distributed network without a fusion center, represented by a connected graph among multiple agent nodes. Each agent node is equipped with a sensor, and   uses a Kalman consensus information filter (KCIF) to track a discrete time global process with linear dynamics and additive Gaussian noise. The state estimate of  the global process at any sensor is computed from the local observation history and the information received by that agent node from its neighbors. At an unknown time, an attacker starts altering the local observation of one agent node. In the Bayesian setting where there is a known prior distribution of the attack beginning instant, we formulate a Bayesian quickest change detection (QCD) problem for FDI detection in order to minimize the mean detection delay subject to a false alarm probability constraint. While it is well-known that the optimal Bayesian QCD rule involves checking the Shriyaev's statistic against a threshold, we  demonstrate how to compute the Shriyaev's statistic at each node in a recursive fashion given our non-i.i.d. observations. Next, we consider non-Bayesian QCD where the attack begins at an arbitrary and unknown time, and the detector seeks to minimize the worst case detection delay subject to a constraint on the mean time to false alarm and probability of misidentification. We use the multiple hypothesis sequential probability ratio test for attack detection and identification at each sensor. For unknown attack strategy, we use the window-limited generalized likelihood ratio (WL-GLR) algorithm to solve the QCD problem. Numerical results demonstrate the performances and trade-offs of the proposed algorithms.
\end{abstract}

\section{Introduction}\label{section:introduction}
Cyber physical systems (CPS) integrate the physical processes and systems with various operations in the cyber world such as sensing, communication, control and computation  \cite{poovendran2011special}. The rapid growth of CPS and its safety-critical applications have resulted in a surge of interest in CPS security in recent years\cite{sandberg2015cyberphysical}. Such applications cover a range of systems such as autonomous vehicles, smart power grids, smart home and automated industries. Cyber attack on such systems   result in malfunctioning and even accidents.  

Attacks on CPS can be mainly categorized into two types: (i)  denial-of-service (DoS) attacks, and (ii) deception attacks. In DoS attack, the attacker makes a particular service or resource such as bandwidth inaccessible to the system. However, our current study of FDI attacks falls under the category of deception attacks wherein the adversary injects  malicious  data into the system, thereby altering the state variables and rendering estimation of these state variables prone to errors. In particular, we develop the quickest detection algorithms against FDI on a network performing distributed tracking of a global state at each node.

There have been significant activities in recent years in the field of FDI in CPS. Early research works on deception attacks in CPS were introduced in \cite{liu2011false}. The attack design and characterization has been carried out in a number of papers that work on optimal attack strategies for maximizing error in estimation. Such FDI attack designs are introduced in \cite{guo2016optimal},\cite{chen2017optimal} and are undetectable to the most common detectors in use.
In \cite{an2017data}, the authors have shown how the attacker can bypass the detector at the estimator side without having the exact knowledge of the system parameters. In \cite{li2019optimal}, arbitrary mean Gaussian distributed injections are considered, along with analysis of the system performance by looking at the statistical characteristics of the measurement innovation. The developed attack scheme achieves the largest remote estimation error and guarantees stealth from the detector. In \cite{moradi2019coordinated}, consensus-based distributed Kalman filtering subject to data-falsification attack is investigated, where the Byzantine agents share manipulated data with their neighboring agents. The design of attack is such that it maximizes the network wide estimation error by adding Gaussian noise. The paper \cite{choraria2022design} proposes a linear attack design in a distributed scenario wherein the estimates of nodes are steered to a desired target with a constraint on the attack detection probability. 

Recent literature has shown that the commonly used windowed $\chi^2$-detector can be circumvented by an attacker if it maintains the same statistics of the innovation sequence before and after the attack   \cite{guo2016optimal}. Naturally, researchers has come up with various  detection approaches for {\em centralized} detection of FDI; e.g., Gaussian Mixture Model (GMM) approach  \cite{guo2018secure},  exploiting correlation between safe (trusted) and unsafe sensors  \cite{li2017detection},  FDI detection in randomized gossip-based sensor networks \cite{gentz2016data}, and  checking the anomaly in the estimates returned by various sensor subsets \cite{chattopadhyay2019security}. On the other hand, there has been several works on FDI detection and mitigation in a  networked system. For example,  signal detection subject to Byzantine attacks in a distributed network has   been investigated in \cite{kailkhura2016data} where a consensus algorithm is employed to enable each node in the network to locally calculate a global decision statistic. The study in \cite{guan2017distributed} focuses on joint distributed attack detection and   secure estimation under physical and cyber attacks, and proposes a  residue based threshold detection scheme. A KL-divergence based detector aided with a watermarking strategy using pseudo-random numbers has been proposed in \cite{zhou2022watermarking} to detect a stealthy attack. A distributed attack detection problem    has been studied in \cite{li2019distributed} for sensors that share quantized local statistics to achieve optimal decision-making with minimal communication costs. The paper  \cite{hao2020consensus} has considered  Kalman consensus filtering in a distributed setting,   and used $\chi^2$ detector for attack detection at each node.

However, in real-time CPS operation, the importance of prompt and accurate detection of cyber attacks is   paramount. The  quickest change detection (QCD \cite{poor2008quickest,tartakovsky2019sequential,tartakovsky2014sequential}) framework can model this problem.  In this setup, the detector sequentially makes observations, and decides at each time whether to stop and declare that a change has occurred to the observation statistics or to continue gathering the measurements;   the goal    is to minimize detection delay while meeting a constraint on the false alarm rate.  In fact,  decentralized QCD has also been extensively studied   \cite{veeravalli2001decentralized,mei2005information, moustakides2006decentralized,banerjee2016decentralized,huang2021asymptotic}, however here a central fusion center gathers measurements from the nodes of a network and makes a decision at each time. Obviously, QCD has been explored in a number of prior works to detect  FDI in centralized setting. The paper \cite{nath2019quickest} has investigated QCD of FDI in smart grids, by utilizing a normalized Rao-CUSUM detector. The authors of \cite{gupta2021quickest} have investigated   Bayesian QCD using Markov decision process (MDP) formulation.
The paper \cite{kurt2018distributed} has performed both centralized and decentralized QCD   of FDI by using the generalized CUSUM algorithm, where the decentralized setting involves local centers transmitting quantized messages to a global center using level-crossing sampling with hysteresis. On the other hand, a    distributed   change detection algorithm in the non-Bayesian setting has been studied in \cite{hsu2023efficient}, where each sensor uses its own observation and the quantized messages from the neighbors to make its decision based on a CUSUM-like statistic. However, unlike our current work, the paper \cite{hsu2023efficient} neither considers QCD of FDI nor considers a distributed Kalman-consensus filter (KCF \cite{olfati2007distributed}) based distributed tracking of a global state.  In fact, QCD of FDI in a distributed system using KCF for distributed tracking of a global state has not been explored in the literature, which motivates our work in this paper.

In this paper, we make the following contributions:
\begin{enumerate}
    \item For a distributed state tracking system using Kalman consensus information filter (KCIF), we develop local quickest detection algorithms for FDI detection. Our algorithms can detect FDI from the change in statistics of  the information flowing through the network. The algorithms also allow a node to detect FDI at some other node, and also to identify the node under attack.
    \item We formulate the QCD problem in a Bayesian setting with a known prior on the attack beginning instant. While many studies utilize Shriyaev's test for change detection, the requirement for a local decision making at a node and the exchange of  local estimates of the process among neighboring nodes in KCIF motivate us to  use the non-i.i.d. consensus estimates  as the data points for attack detection.  Despite the complex dependencies due to the distributed system and the usage of KCIF, we provide  a computationally efficient, recursive update rule for the detection statistic, i.e., the posterior belief at any node on the attack having already started at another node.  
    \item In case of non-Bayesian detection where the attack starts at an arbitrary time, we adapt  quickest detection and isolation algorithms from the literature to detect  FDI in this distributed setting, for known and unknown attack strategies. Interestingly, the recursive update rules developed for the Bayesian QCD problem are again used here to efficiently compute the test statistic.
    \end{enumerate}

The rest of this paper is organized as follows. 
Section \ref{section:System Model} introduces the system model.  In Section \ref{section:Dynamics of estimate}, we derive the recursive update rules for conditional densities of consensus estimates and the observation required for the detection. Section \ref{section:Bayesian Detection} deals with Bayesian QCD of FDI in the network, while   Section \ref{section:Non-Bayesian QCD}  deals with Non-Bayesian QCD. We present the numerical results in Section \ref{section:Numerical Results}. Section \ref{section:Conclusion} concludes the paper.

\section{System model}\label{section:System Model}
In this paper, bold capital letters, bold small letters, and capital letters with calligraphic font will denote matrices, vectors and sets, respectively.

\begin{figure}[!htb]
  \includegraphics[scale=1.7]{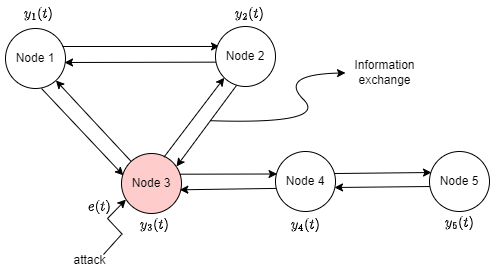}
  \caption{False data injection attack in a distributed setting.}
  \label{fig:topology}
\end{figure}
\subsection{Sensing and  estimation}\label{subsection:sensing and estimation}
We consider a distributed system with $N$ sensors, also called nodes, denoted by $\mathcal{N}\doteq\{1,2,\cdots,N\}$. The set of neighboring nodes of node~$i$ is denoted by $\mathcal{N}_i$. Let  $\mathcal{N}'_i \doteq \mathcal{N}_i \cup \{i\}$ and $N_i \doteq |\mathcal{N}_i|$ be the cardinality of the set $\mathcal{N}_i$ . The sensors observe a single discrete time process $\{\bm{x}(t)\}_{t \geq 1}$, with a Kalman consensus filter employed at each sensor node for estimation of $\bm{x}(t)$ (Figure \ref{fig:topology}).
The system evolves as a linear  process with Gaussian noise:
\begin{equation}\label{eqn:process-equation}
	\bm{x}(t) = \bm{Ax}(t-1) + \bm{w}(t-1) 
\end{equation}
where $\bm{x}(t)\in\mathbb{R}^{p \times 1}$ is the state at time $t$,
$\bm{w}(t) \sim \mathcal{N}(\bm{0}, \bm{Q})$ is the zero-mean Gaussian process noise i.i.d. across time and $\bm{A} \in \mathbb{R}^{p \times p}$ is the process matrix or the state transition matrix.

At time $t$,  each sensor~$i$ makes an observation
\begin{equation}\label{eqn:observation-equation}
    \bm{y}_i(t) = \bm{C}_i \bm{x}(t) + \bm{v}_i(t) ,
\end{equation}

\noindent where $\bm{y}_i(t)\in\mathbb{R}^{q \times 1}$ denotes the vector of system measurements, and $\bm{C}_i \in \mathbb{R}^{q \times p}$ is the observation matrix. Also, $\bm{v}_i(t) \sim \mathcal{N}(\bm{0}, \bm{R}_i)$ is the zero-mean Gaussian measurement noise of sensor $i$ with covariance $\bm{R}_i$,  and is independent across sensors and i.i.d. across $t$. The pair $(\bm{A}, \bm{Q}^{\frac{1}{2}})$ is assumed to be stabilizable, and the pair $(\bm{A},\bm{C}_i)$ is assumed to be observable for each $1 \leq  i \leq N$.
The initial state, $\bm{x}(0)$ is taken to be zero mean Gaussian with covariance matrix $\bm{P}(0)$, independent of the noise sequences.

\begin{figure*}[htb]
\hrule

Kalman consensus information filtering equations at sensor $i$
\footnotesize
\begin{IEEEeqnarray}{rCl}\label{eqn:KCF-estimate}
     \bm{u}_{i}(t) &=& \bm{C}_i'\bm{R}_i^{-1}\bm{y}_i(t) \qquad \bm{U}_i = \bm{C}_i'\bm{R}_i^{-1}\bm{C}_i \nonumber\\
     \bm{\phi}_i(t) &=& \sum_{j\in \mathcal{N}'_i} \bm{u}_{j}(t) \qquad \bm{S}_i =\sum_{j\in \mathcal{N}'_i} \bm{U}_j \nonumber\\
     \hat{\bm{x}}_{i}(t) &=& \bm{A}\hat{\bm{x}}_{i}(t-1) + \bm{M}_i(t)(\bm{\phi}_i(t) - \bm{S}_i\bm{A} \hat{\bm{x}}_{i}(t-1)) +\gamma\bm{P}_i(t)\bm{A} \sum_{j \in \mathcal{N}_i}(\hat{\bm{x}}_{j}(t-1) - \hat{\bm{x}}_{i}(t-1)) \nonumber\\
     &=& \bm{M}_i(t)\bm{S}_i\bm{A}\bm{x}(t-1) + (\bm{A} - \bm{M}_i(t)\bm{S}_i\bm{A})\hat{\bm{x}}_{i}(t-1)+ \gamma\bm{P}_i(t)\bm{A} \sum_{j \in \mathcal{N}_i}(\hat{\bm{x}}_{j}(t-1) - \hat{\bm{x}}_{i}(t-1)) \nonumber\\
     &\qquad+& \bm{M}_i(t)\sum_{j\in \mathcal{N}'_i}\bm{C}_j'\bm{R}_j^{-1}\bm{v}_j(t) + \bm{M}_i(t)\bm{S}_i\bm{w}(t-1) \nonumber\\
     \bm{M}_i(t) &=& (\bm{P}_i(t)^{-1} + \bm{S}_i)^{-1} \nonumber\\ 
    \bm{P}_i(t+1) &=& \bm{A}\bm{M}_i(t)\bm{A}' +\bm{Q} \nonumber\\
    \gamma &=& \frac{\epsilon}{(\Vert \bm{P}_i(t)\Vert_F+1)}
\end{IEEEeqnarray}
$\gamma\bm{P}_i(t)$ represents the appropriate consensus gain for a stable filter\cite{olfati2009kalman}. \\ $\Vert \cdot \Vert_F$ represents the Frobenius norm of a matrix and $\epsilon>0$ is a small constant.
\normalsize
\hrule
\end{figure*}

In this paper, we employ the Kalman consensus information filter (KCIF \cite{olfati2009kalman}) for distributed estimation. The estimate of the process at time $t$ at sensor $i$ is denoted as $\hat{\bm{x}}_{i}(t)$. Also, we denote the predicted estimate as $\bar{\bm{x}}_{i}(t)=\bm{A}\hat{\bm{x}}_{i}(t-1)$. Note that, $\hat{\bm{x}}_{i}(t)$ is an unbiased estimate of the process $\bm{x}(t)$ and is Gaussian distributed. We define the vertical concatenation of the estimates declared by all neighbors of sensor $i$ at time $t$ as $\hat{\bm{x}}_{\mathcal{N}_i}(t)$. Let us define $\mathcal{I}_{i,\mathcal{N}_i}(t, t', t'') = \{\hat{\bm{x}}_{i}(1),\ldots,\hat{\bm{x}}_{i}(t),\hat{\bm{x}}_{\mathcal{N}_i}(1),\ldots, \hat{\bm{x}}_{\mathcal{N}_i}(t'-1),\bm{y}_i(t'')\}$ as the available information at sensor $i$ at time $t$, which comprises the sensor's estimates up to time $t$, the estimates from the neighboring sensors $\mathcal{N}_i$ up to time $t'-1$ and the measurement at time $t''$. Let $\bm{\zeta}_i(t)\doteq \hat{\bm{x}}_{i}(t) - \bm{x}(t)$ be the posterior estimation error and $\bar{\bm{\zeta}}_i(t)\doteq\bar{\bm{x}}_{i}(t) - \bm{x}(t)$ be the prediction error. We define prediction error covariance and estimation error covariance as $\bm{P}_i(t) = \mathbb{E}[\bar{\bm{\zeta}}_i(t)\bar{\bm{\zeta}}'_i(t)]$ and $\bm{M}_i(t)=\mathbb{E}[\bm{\zeta}_i(t)\bm{\zeta}'_i(t)]$, respectively.

The detailed update equations of the KCIF are given in \eqref{eqn:KCF-estimate}. 
At time $t$, each sensor $i \in \mathcal{N}$ receives the message $\{\hat{\bm{x}}_{\mathcal{N}_i}(t-1), \bm{u}_{\mathcal{N}_i}(t), \bm{U}_{\mathcal{N}_i}\}$ from its neighbors, as described in \eqref{eqn:KCF-estimate}. Based on this, sensor $i$ calculates the ``fused information matrix'' $\bm{S}_i$ and ``fused weighted measurement vector'' $\bm{\phi}_i(t)$. Subsequently, sensor $i$ computes the final consensus estimate $\hat{\bm{x}}_{i}(t)$.

\subsection{FDI Attack Model}\label{subsection:FDI-attack}
We assume that at most one sensor can be under attack, though the theory developed in this paper can be extended to address the case where more than one sensors are under attack. The measurement made by sensor $i$ at time $t$ under FDI attack is represented as
\begin{equation}\label{eqn:observation-under-attack}
	\tilde{\bm{y}}_{i}(t) = \bm{C}_i \bm{x}(t) + \bm{v}_{i}(t) + \bm{e}_{i}(t) ,
\end{equation}
where $\bm{e}_i(t) \sim \mathcal{N}(0,\bm{\Sigma})$ is the Gaussian distributed  error injected by the attacker at the $i^{th}$ sensor at time $t$, and the covariance matrix $\bm{\Sigma}$ is known to the defender.  We assume that  the attacker initiates the attack at an unknown time $\tau$, and the detector at each node seeks to detect the attack.

Let the   innovation be denoted by $\bm{z}_i(t) \doteq \bm{y}_i(t) - \bm{C}_i\bm{A}\bm{\hat{x}}_i(t-1)$. In a remote estimation setting, a commonly used FDI detector is the windowed $\chi^2$ detector. At each time $t$, it tests whether  $\sum_{j=t-J+1}^{t}\bm{z}_i(j)'cov(\bm{z}_i(j))^{-1}\bm{z}_i(j) \geq \eta$, where $\eta>0$ is a threshold to control the false alarm rate, and $J$ is the pre-specified window size.  If this condition is met, at time~$t$, the detector raises an alarm. 

Even though there has been work done on quickest change detection, most of these either focus on the centralized setting or a decentralized case with a fusion center. In the latter case, the sensors send their local information to a central fusion center which carries out the detection of any attack. In this work, we assume that each detector has a  sensor. Due to distributed setting, information broadcasting and consensus results in error propagation to unattacked sensors over time, and therefore a change can be detected at any benign sensor as well.

\section{Dynamics of the KCIF estimate}\label{section:Dynamics of estimate}
 
In this section, we  consider the KCIF update rule for $\hat{\bm{x}}_{i}(t)$  in \eqref{eqn:KCF-estimate}, and seek to compute its distribution conditioned on the information $\mathcal{I}_{i,\mathcal{N}_i}(t-1,t,t)$ available at sensor~$i$. 
\subsection{Conditional covariance of $\hat{\bm{x}}_i(t)$}\label{subsection:conditional-covariance-of-estimate}
Let us denote the conditional covariance of the estimate $\hat{\bm{x}}_{i}(t)$ by $\bm{\hat{P}}_i(t)$. Let $
\hat{\bm{r}}_i(t) \doteq  
\begin{bmatrix} 
\hat{\bm{x}}'_i(t-1) & \hat{\bm{x}}'_{\mathcal{N}_i}(t-1)  & \bm{y}'_i(t) 
\end{bmatrix}^{'}
$.
\noindent We approximate $\bm{\hat{P}}_i(t)$ as 
\begin{IEEEeqnarray}{rCl}\label{eqn:approximate-conditional-covariance-of-estimate}
    \bm{\hat{P}}_i(t) &=& cov[\hat{\bm{x}}_{i}(t) | \mathcal{I}_{i,\mathcal{N}_i}(t-1,t,t)] \nonumber\\
    &\approx& cov[\hat{\bm{x}}_{i}(t)|\hat{\bm{x}}_{i}(t-1),\hat{\bm{x}}_{\mathcal{N}_i}(t-1),\bm{y}_{i}(t)] \nonumber\\
    &=& cov[\hat{\bm{x}}_{i}(t)|\hat{\bm{r}}_i(t)]
\end{IEEEeqnarray}
where we assume that the current estimate is conditionally independent of all past information given $\hat{\bm{r}}_i(t)$.

\noindent For joint Gaussian random variables, we have,
\begin{IEEEeqnarray}{rCl}\label{eqn:joint_gaussian_rvs}
    &&cov[\hat{\bm{x}}_{i}(t)|\hat{\bm{r}}_i(t)] \nonumber\\
    &=& cov[\hat{\bm{x}}_{i}(t)] - cov[\hat{\bm{x}}_{i}(t),\hat{\bm{r}}_i(t)] cov[\hat{\bm{r}}_i(t)]^{-1}cov[\hat{\bm{r}}_i(t),\hat{\bm{x}}_{i}(t)] \IEEEeqnarraynumspace
\end{IEEEeqnarray}

\subsubsection{Some useful updates}

To calculate the terms in \eqref{eqn:joint_gaussian_rvs}, we define the following terms for simplification: 
\begin{IEEEeqnarray*}{rCl}
    \bm{D}_{i,t} &=& \bm{A} - \bm{M}_i(t)\bm{S}_i\bm{A} -\gamma\bm{P}_i(t)\bm{A}N_i \\ 
    \bm{G}_{i,t} &=& \bm{M}_i(t)\bm{S}_i\bm{A} \\
    \bm{F}_{i,t} &=& \bm{P}_i(t)\bm{A} \\ 
    \bm{B}(t) &=& cov[\bm{x}(t)] \\
    \bm{L}_i(t) &=& cov[\hat{\bm{x}}_i(t)] \\ 
    \bm{H}_i(t) &=& cov[\hat{\bm{x}}_i(t),\bm{x}(t)] \\
    \bm{T}_{i,j}(t) &=& cov[\hat{\bm{x}}_i(t),\hat{\bm{x}}_j(t)], \forall j \in \mathcal{N}_i
\end{IEEEeqnarray*}

\noindent The KCIF estimate in \eqref{eqn:KCF-estimate} can then be rewritten as 
\begin{IEEEeqnarray}{rCl}\label{eqn:rewritten_estimate}
    \hat{\bm{x}}_{i}(t) &=& \bm{G}_{i,t}\bm{x}(t-1)+\bm{D}_{i,t}\hat{\bm{x}}_{i}(t-1)+\gamma\bm{F}_{i,t} \sum_{j \in \mathcal{N}_i}\hat{\bm{x}}_{j}(t-1) \nonumber\\
    &+& \bm{M}_i(t)\sum_{j\in \mathcal{N}'_i }\bm{C}_j'\bm{R}_j^{-1}\bm{v}_j(t) + \bm{M}_i(t)\bm{S}_i\bm{w}(t-1)
\end{IEEEeqnarray}
\noindent The covariance of the second last term in \eqref{eqn:rewritten_estimate} is:
\begin{IEEEeqnarray}{rCl}
    && cov\left(\bm{M}_i(t)\sum_{j\in \mathcal{N}'_i}\bm{C}_j'\bm{R}_j^{-1}\bm{v}_j(t)\right) \nonumber\\
    &=& \bm{M}_i(t)\left(\sum_{j\in \mathcal{N}'_i}\bm{C}_j'\bm{R}_j^{-1}\bm{R}_j(\bm{R}_j^{-1})'\bm{C}_j\right)\bm{M}_i'(t) \nonumber\\
    &=&\bm{M}_i(t)\underbrace{\left(\sum_{j\in \mathcal{N}'_i}\bm{C}_j'(\bm{R}_j^{-1})'\bm{C}_j \right)}_{\doteq \bm{S}_i'}\bm{M}_i'(t) 
\end{IEEEeqnarray}

\subsubsection{Calculating first term in \eqref{eqn:joint_gaussian_rvs}}
The first term in \eqref{eqn:joint_gaussian_rvs} is the unconditional covariance of $\hat{\bm{x}}_i(t)$. Using \eqref{eqn:rewritten_estimate}, we have: 
\begin{IEEEeqnarray}{rCl}\label{eqn:unconditional-covariance-of-estimate}
    \bm{L}_i(t) &=& \bm{G}_{i,t} \bm{B}(t-1)\bm{G}'_{i,t}+ \bm{G}_{i,t}\bm{H}'_i(t-1) \bm{D}'_{i,t} \nonumber\\
    &+& \gamma \bm{G}_{i,t} \sum_{j\in \mathcal{N}_i}\bm{H}'_j(t-1)\bm{F}'_{i,t} + \bm{D}_{i,t}\bm{H}_i(t-1) \bm{G}'_{i,t} \nonumber\\
    &+& \bm{D}_{i,t} \bm{L}_i(t-1)\bm{D}'_{i,t} + \gamma\bm{D}_{i,t} \sum_{j\in \mathcal{N}_i}\bm{T}_{i,j}(t-1)\bm{F}'_{i,t} \nonumber\\
    &+& \scalemath{0.95}{\gamma\bm{F}_{i,t} \sum_{j\in \mathcal{N}_i}\bm{H}_j(t-1)\bm{G}'_{i,t} + \gamma\bm{F}_{i,t} \sum_{j\in \mathcal{N}_i}\bm{T}'_{i,j}(t-1)\bm{D}'_{i,t}}\nonumber \\ 
    &+&\scalemath{0.9}{\gamma^2 \bm{F}_{i,t}\sum_{j\in \mathcal{N}_i}\bm{L}_j(t-1)\bm{F}'_{i,t} + \gamma^2\bm{F}_{i,t} \sum_{\substack{j,k\in \mathcal{N}_i \\ j \neq k}} \bm{T}_{j,k}(t-1)\bm{F}'_{i,t}} \nonumber\\
    &+&\bm{M}_i(t)\bm{S}'_i\bm{M}'_i(t) + \bm{M}_i(t)\bm{S}_i \bm{Q}\bm{S}'_i\bm{M}'_i(t)
\end{IEEEeqnarray}

The above update requires calculating the covariance between $\hat{\bm{x}}_i(t)$ and $\bm{x}(t)$, i.e., $\bm{H}_i(t)$ and the cross covariances between neighboring estimates, i.e., $\bm{T}_{i,j}(t)$.

\noindent With $\bm{x}(t)=\bm{A}\bm{x}(t-1) + \bm{w}(t-1)$,
\begin{IEEEeqnarray}{rCl}\label{eqn:cross-cov-of-estimate-and-process}
    \bm{H}_i(t) &=& cov[\hat{\bm{x}}_i(t),\bm{x}(t)] \nonumber\\
    &=& \bm{G}_{i,t}\bm{B}(t-1)\bm{A}'+\bm{D}_{i,t} \bm{H}_i(t-1) \bm{A}' \nonumber\\
    &+& \gamma \bm{F}_{i,t}\sum_{j\in \mathcal{N}_i}\bm{H}_j(t-1) \bm{A}'  +  \bm{M}_i(t)\bm{S}_i \bm{Q}
\end{IEEEeqnarray}     

At sensor $i$, we assume $\bm{L}_l(t), \bm{H}_l(t), \bm{T}_{i,l}(t), \bm{T}_{l,m}(t)$ to be zero, where $l,m \notin \mathcal{N}_i, \forall t \geq 1$. Also, at time $t$, sensor $i$ has access to $\{\bm{H}_j(t'), \bm{L}_j(t'):  j \in \mathcal{N}_i, 1 \leq t' \leq t-1\}$ due to local information sharing at each time step. With this assumption,  we proceed to find $\bm{T}_{i,j}(t)$ and $\bm{T}_{j,k}(t)$ at sensor $i$ using \eqref{eqn:rewritten_estimate}, where $j,k \in \mathcal{N}_i, j\neq k$.

\begin{IEEEeqnarray}{rCl}\label{eqn:cross-cov-between-estimate-at-sensor i-and-sensor j}
    \bm{T}_{i,j}(t) &=& \bm{G}_{i,t} \bm{B}(t-1)\bm{G}'_{j,t} + \bm{G}_{i,t} \bm{H}'_j(t-1)\bm{D}'_{j,t} \nonumber\\
    &+& \gamma\bm{G}_{i,t} \sum_{r \in \mathcal{N}_j\cap\mathcal{N}'_i}\bm{H}'_r(t-1)\bm{F}'_{j,t} + \bm{D}_{i,t} \bm{H}_i(t-1)\bm{G}'_{j,t} \nonumber\\ 
    &+&  \scalemath{0.98}{\bm{D}_{i,t}\bm{T}_{i,j}(t-1)\bm{D}'_{j,t} + \gamma\bm{D}_{i,t}\sum_{r\in \mathcal{N}_j\cap \mathcal{N}'_i}\bm{T}_{i,r}(t-1)\bm{F}'_{j,t}} \nonumber\\
    &+& \scalemath{0.95}{\gamma\bm{F}_{i,t} \sum_{r\in \mathcal{N}_i}\bm{H}_r(t-1)\bm{G}'_{j,t} +  \gamma\bm{F}_{i,t} \sum_{r\in \mathcal{N}_i} \bm{T}_{r,j}(t-1)\bm{D}'_{j,t}} \nonumber\\
    &+& \gamma^2 \bm{F}_{i,t} \sum_{\substack{r\in\mathcal{N}_i \\s\in \mathcal{N}_j \cap\mathcal{N}'_i}}\bm{T}_{r,s}(t-1) \bm{F}'_{j,t} \nonumber\\
    &+&  \scalemath{0.96}{\bm{M}_i(t)\sum_{r\in\mathcal {N}'_j \cap\mathcal{N}'_i}\bm{U}'_r \bm{M}'_j(t) + \bm{M}_i(t)\bm{S}_i\bm{Q}\bm{S}'_j\bm{M}'_j(t)}\IEEEeqnarraynumspace
\end{IEEEeqnarray}
Note that if $r=s, \bm{T}_{r,s}(t) = \bm{L}_r(t)$.

\noindent Now
\begin{IEEEeqnarray}{rCl}\label{eqn:cross-cov-between-estimates-of-nieghbours-of-sensor i}
    \bm{T}_{j,k}(t) &=& \bm{G}_{j,t} \bm{B}(t-1)\bm{G}'_{k,t} + \bm{G}_{j,t} \bm{H}'_k(t-1)\bm{D}'_{k,t} \nonumber\\
    &+& \gamma\bm{G}_{j,t} \sum_{r \in \mathcal{N}_k\cap \mathcal{N}'_i}\bm{H}'_r(t-1)\bm{F}'_{k,t} + \bm{D}_{j,t} \bm{H}_j(t-1)\bm{G}'_{k,t} \nonumber\\ 
    &+&  \scalemath{0.98}{\bm{D}_{j,t}\bm{T}_{j,k}(t-1)\bm{D}'_{k,t} + \gamma\bm{D}_{j,t}\sum_{r\in \mathcal{N}_k\cap \mathcal{N}'_i}\bm{T}_{j,r}(t-1)\bm{F}'_{k,t}} \nonumber\\
    &+& \scalemath{0.85}{\gamma\bm{F}_{j,t} \sum_{r\in \mathcal{N}_j\cap \mathcal{N}'_i}\bm{H}_r(t-1)\bm{G}'_{k,t} +  \gamma\bm{F}_{j,t} \sum_{r\in \mathcal{N}_j\cap \mathcal{N}'_i} \bm{T}_{r,k}(t-1)\bm{D}'_{k,t}} \nonumber\\
    &+& \gamma^2 \bm{F}_{j,t} \sum_{\substack{r\in\mathcal{N}_j \cap\mathcal{N}'_i\\s\in \mathcal{N}_k  \cap\mathcal{N}'_i}} \bm{T}_{r,s}(t-1) \bm{F}'_{k,t} \nonumber\\ 
    &+& \scalemath{0.98}{\bm{M}_j(t)\sum_{r\in\mathcal{M}}\bm{U}'_r \bm{M}'_k(t)+\bm{M}_j(t)\bm{S}_j\bm{Q}\bm{S}'_k\bm{M}'_k(t)}
\end{IEEEeqnarray}
where $\mathcal{M} = \mathcal {N}'_j \cap\mathcal{N}'_k\cap\mathcal{N}'_i$.

We can use the results in \eqref{eqn:cross-cov-of-estimate-and-process}, \eqref{eqn:cross-cov-between-estimate-at-sensor i-and-sensor j} and \eqref{eqn:cross-cov-between-estimates-of-nieghbours-of-sensor i}, to obtain the final update for $\bm{L}_i(t)$ in \eqref{eqn:unconditional-covariance-of-estimate}.


\subsubsection{Calculating the remaining terms in \eqref{eqn:joint_gaussian_rvs}}
\begin{IEEEeqnarray}{rCl}\label{eqn:cross-cov-between-estimate at i-and-information at i}
    && cov(\hat{\bm{x}}_i(t),\hat{\bm{r}}_i(t))\nonumber\\ &=&\mathbb{E}\left(\hat{\bm{x}}_i(t) 
    \begin{bmatrix} 
    \hat{\bm{x}}'_i(t-1) & \hat{\bm{x}}'_{\mathcal{N}_i}(t-1)  & \bm{y}'_i(t) 
    \end{bmatrix} 
    \right)\nonumber\\
    &=&
    \begin{bmatrix} 
    cov(\hat{\bm{x}}_{i}(t), \hat{\bm{x}}_i(t-1)) \\
    cov(\hat{\bm{x}}_{i}(t), \hat{\bm{x}}_{\mathcal{N}_i}(t-1)) \\
    cov(\hat{\bm{x}}_{i}(t),\bm{y}_i(t)) 
    \end{bmatrix}^{'}
\end{IEEEeqnarray}

\noindent Using definition of $\hat{\bm{x}}_{i}(t)$ in \eqref{eqn:rewritten_estimate}, we have
\begin{IEEEeqnarray}{rCl} \label{eqn:cross-cov-between-unit delay-estimates at sensor i}
    cov(\hat{\bm{x}}_{i}(t), \hat{\bm{x}}_i(t-1)) &=& \bm{G}_{i,t}\bm{H}'_i(t-1) + \bm{D}_{i,t} \bm{L}_i(t-1) \nonumber \\
    &+& \gamma\bm{F}_{i,t} \sum_{j\in
    \mathcal{N}_i}\bm{T}'_{i,j}(t-1)
\end{IEEEeqnarray}
Again from \eqref{eqn:rewritten_estimate}, for all $j \in \mathcal{N}_i$,  
\begin{IEEEeqnarray}{rCl}\label{eqn:cross-cov-between-estimate at i-and-unit delayed-estimate-of-neighbor}
    cov(\hat{\bm{x}}_{i}(t), \hat{\bm{x}}_j(t-1)) &=& \bm{G}_{i,t}\bm{H}'_j(t-1) + \bm{D}_{i,t} \bm{T}_{i,j}(t-1) \nonumber\\
    &+& \gamma\bm{F}_{i,t} \sum_{r\in\mathcal{N}_i}\bm{T}_{r,j}(t-1)
\end{IEEEeqnarray}

\noindent Using $\bm{y}_i(t) = \bm{C}_i\bm{x}(t) + \bm{v}_i(t)$ and the definition of $\hat{\bm{x}}_i(t)$ in \eqref{eqn:rewritten_estimate}, we have,
\begin{IEEEeqnarray}{rCl}\label{eqn:cross-cov-between-estimate at i-and-observation at i}
    cov(\hat{\bm{x}}_{i}(t),\bm{y}_i(t)) &=&cov(\hat{\bm{x}}_{i}(t), \bm{C}_i\bm{x}(t) + \bm{v}_i(t)) \nonumber\\
    &=&\bm{H}_i(t)\bm{C}'_i  + \bm{M}_i(t)\bm{C}'_i\bm{R}_i^{-1}\bm{R}_i \nonumber\\
    &=&(\bm{H}_i(t) + \bm{M}_i(t))\bm{C}'_i
\end{IEEEeqnarray}
Using \eqref{eqn:cross-cov-between-unit delay-estimates at sensor i}, \eqref{eqn:cross-cov-between-estimate at i-and-unit delayed-estimate-of-neighbor} and \eqref{eqn:cross-cov-between-estimate at i-and-observation at i}, we can compute $cov(\hat{\bm{x}}_i(t),\hat{\bm{r}}_i(t))$ in \eqref{eqn:cross-cov-between-estimate at i-and-information at i}. 

\noindent Lastly, we are left with 
\begin{IEEEeqnarray}{rCl}\label{eqn:covariance-of-information at i}
    && cov(\hat{\bm{r}}_i(t)) \nonumber\\
    &=& \mathbb{E}\left(\begin{pmatrix} 
    \hat{\bm{x}}_i(t-1) \\ \hat{\bm{x}}_{\mathcal{N}_i}(t-1) \\ \bm{y}_i(t)  
    \end{pmatrix} 
    \begin{pmatrix} \hat{\bm{x}}_i(t-1) \\ \hat{\bm{x}}_{\mathcal{N}_i}(t-1) \\ \bm{y}_i(t)  
    \end{pmatrix}^{'} \right) \nonumber\\
    &=&\scalemath{0.95}{\begin{bmatrix}
    \bm{L}_i(t-1) & \bm{T}_{i,{\mathcal{N}_i}}(t-1)&\bm{H}_i(t-1)\bm{A}'\bm{C}'_i \\   \bm{T}'_{i,{\mathcal{N}_i}}(t-1) & \bm{T}_{\mathcal{N}_i,\mathcal{N}_i}(t-1) & \bm{H}_{\mathcal{N}_i}(t-1)\bm{A}'\bm{C}'_i \\ \bm{C}_i \bm{A} \bm{H}'_i(t-1) & \bm{C}_i \bm{A} \bm{H}'_{\mathcal{N}_i}(t-1) & \bm{C}_i \bm{B}(t) \bm{C}'_i + \bm{R}_i
    \end{bmatrix}} \IEEEeqnarraynumspace
\end{IEEEeqnarray}

where 
\begin{IEEEeqnarray*}{rCl}
    &\bm{T}_{i,{\mathcal{N}_i}}(t-1) &= \begin{bmatrix} 
    \bm{T}'_{i,j_1}(t-1) \\ \vdots \\ \bm{T}'_{i,j_{N_i}}(t-1)  
    \end{bmatrix}^{'} \\
    &\bm{T}_{\mathcal{N}_i,\mathcal{N}_i}(t-1) &= \begin{bmatrix}
        \bm{T}_{j_1,j_1}(t-1) & \ldots & \bm{T}_{j_1,j_{N_i}}(t-1) \\ \vdots & \ddots & \vdots \\
        \bm{T}_{j_{N_i},j_1}(t-1) & \ldots & \bm{T}_{j_{N_i},j_{N_i}}(t-1)
    \end{bmatrix} \\
    & \bm{H}_{\mathcal{N}_i}(t-1) &= \begin{bmatrix} 
    \bm{H}'_{j_1}(t-1) & \ldots & \bm{H}'_{j_{N_i}}(t-1)
    \end{bmatrix}^{'}
\end{IEEEeqnarray*}

where, $\{j_1, \ldots, j_{N_i}\}$ are the neighbors of sensor $i$.

In order to compute $\hat{\bm{P}}_i(t)$ defined in \eqref{eqn:approximate-conditional-covariance-of-estimate}, we use \eqref{eqn:unconditional-covariance-of-estimate}, \eqref{eqn:cross-cov-between-estimate at i-and-information at i} and \eqref{eqn:covariance-of-information at i}.

\subsection{Conditional expectation of $\hat{\bm{x}}_i(t)$}\label{subsection:conditional-expectation-of-estimate}
The conditional expectation of the sensor estimate is 
\begin{IEEEeqnarray*}{rCl}
    \bm{\hat{\mu}}_i(t) &\triangleq& \mathbb{E}[\hat{\bm{x}}_{i}(t) | \mathcal{I}_{i,\mathcal{N}_i}(t-1,t,t)] \\
    &\approxeq& \mathbb{E}[\hat{\bm{x}}_i(t)|\hat{\bm{r}}_i(t)]\\ &=&cov(\hat{\bm{x}}_i(t),\hat{\bm{r}}_i(t))cov(\hat{\bm{r}}_i(t))^{-1} \hat{\bm{r}}_i(t)
\end{IEEEeqnarray*}
\noindent Now to compute $\bm{\hat{\mu}}_i(t)$, we use \eqref{eqn:cross-cov-between-estimate at i-and-information at i} and \eqref{eqn:covariance-of-information at i}.

\subsection{Conditional distribution of neighbor's estimate given information of current sensor}

Here we seek to  calculate the conditional distribution $p(\hat{\bm{x}}_{j}(t-1) | \mathcal{I}_{i,\mathcal{N}_i}(t-1,t-1,t)$.  While calculating this distribution at sensor $i$, we ignore all neighbor's of sensor $j$ other than $i$.

Let us denote the conditional covariance of the estimate $\hat{\bm{x}}_{j}(t-1)$ given $\mathcal{I}_{i,\mathcal{N}_i}(t-1,t-1,t)$ at sensor $i$ by $\bm{\hat{P}}_j^i(t-1)$. As before in section \ref{subsection:conditional-covariance-of-estimate}, we begin by defining $\hat{\bm{r}}_j^i(t)$ as approximation of the information vector, $\mathcal{I}_{i,\mathcal{N}_i}(t-1,t-1,t)$. Let $
\hat{\bm{r}}_j^i(t) \doteq  
\begin{bmatrix} 
\hat{\bm{x}}'_j(t-2) & \hat{\bm{x}}'_i(t-1)  & \bm{y}'_i(t) 
\end{bmatrix}^{'}
$.

We approximate $\bm{\hat{P}}_j^i(t-1)$ as 
\begin{IEEEeqnarray}{rCl}\label{eqn:approximate-conditional-covariance-of-neighbor's-estimate}
    \bm{\hat{P}}_j^i(t-1) &=& cov[\hat{\bm{x}}_{j}(t-1) | \mathcal{I}_{i,\mathcal{N}_i}(t-1,t-1,t)] \nonumber\\
    &\approx& cov[\hat{\bm{x}}_{j}(t-1)|\hat{\bm{x}}_{j}(t-2),\hat{\bm{x}}_i(t-1),\bm{y}_{i}(t)] \nonumber\\
    &=& cov[\hat{\bm{x}}_{j}(t-1)|\hat{\bm{r}}_j^i(t)]
\end{IEEEeqnarray}

Again, we have
\begin{IEEEeqnarray}{rCl}
    &&cov[\hat{\bm{x}}_{j}(t-1)|\hat{\bm{r}}_j^i(t)] \nonumber\\
    &=&cov[\hat{\bm{x}}_{j}(t-1)] \nonumber\\
    &-& cov[\hat{\bm{x}}_{j}(t-1),\hat{\bm{r}}_j^i(t)] cov[\hat{\bm{r}}_j^i(t)]^{-1}cov[\hat{\bm{r}}_j^i(t),\hat{\bm{x}}_{j}(t-1)] \IEEEeqnarraynumspace
\end{IEEEeqnarray}

The first term in above expression is $\bm{L}_j(t-1)$ and requires the knowledge of the matrices $\bm{G}_{j,t-1}, \bm{D}_{j,t-1}, \bm{F}_{j,t-1}, \bm{H}_j(t-2)$ and so on. We assume these parameters are known at sensor $i$.

\noindent Now,
\begin{IEEEeqnarray}{rCl}
    && cov(\hat{\bm{x}}_j(t-1),\hat{\bm{r}}_j^i(t))\nonumber\\ &=&\mathbb{E}\left(\hat{\bm{x}}_j(t-1) 
    \begin{bmatrix} 
    \hat{\bm{x}}'_j(t-2) & \hat{\bm{x}}'_i(t-1)  & \bm{y}'_i(t) 
    \end{bmatrix} 
    \right)\nonumber\\
    &=&
    \begin{bmatrix} 
    cov(\hat{\bm{x}}_{j}(t-1), \hat{\bm{x}}_j(t-2)) \\
    cov(\hat{\bm{x}}_{j}(t-1), \hat{\bm{x}}_i(t-1)) \\
    cov(\hat{\bm{x}}_{j}(t-1),\bm{y}_i(t)) 
    \end{bmatrix}^{'}
\end{IEEEeqnarray}

The first element in above matrix can be calculated in a similar manner as done in \eqref{eqn:cross-cov-between-unit delay-estimates at sensor i}. The second element is simply $\bm{T}'_{i,j}(t-1)$. The last element can be simplified as,
\begin{IEEEeqnarray*}{rCl}
&&cov(\hat{\bm{x}}_{j}(t-1), \bm{C}_i \bm{A}\bm{x}(t-1) + \bm{C}_i \bm{w}(t-1) + \bm{v}_i(t)) \\
&=& \bm{H}_j(t-1)\bm{A}'\bm{C}'_i
\end{IEEEeqnarray*}

\noindent At last, we need $cov[\hat{\bm{r}}_j^i(t)]$ which can be written in a similar way as in \eqref{eqn:covariance-of-information at i}. Using the above equations, we get $\bm{\hat{P}}_j^i(t-1)$ defined in \eqref{eqn:approximate-conditional-covariance-of-neighbor's-estimate}.

The conditional mean of $\hat{\bm{x}}_j(t-1)$ given $\hat{\bm{r}}_j^i(t)$ would be written similarly as in section \ref{subsection:conditional-expectation-of-estimate},
\begin{IEEEeqnarray*}{rCl}
    \bm{\hat{\mu}}_j^i(t) &\triangleq& \mathbb{E}[\hat{\bm{x}}_j(t-1) | \mathcal{I}_{i,\mathcal{N}_i}(t-1,t-1,t)] \\
    &\approxeq& \mathbb{E}[\hat{\bm{x}}_j(t-1)|\hat{\bm{r}}_j^i(t)]\\ &=&cov(\hat{\bm{x}}_j(t-1),\hat{\bm{r}}_j^i(t))cov(\hat{\bm{r}}_j^i(t))^{-1} \hat{\bm{r}}_j^i(t)
\end{IEEEeqnarray*}
where all the terms have been computed beforehand.
\subsection{Conditional distribution of $\bm{y}_i(t)$}\label{subsection:Conditional distribution of observation at i}
In this section we find the conditional distribution of $\bm{y}_i(t)$ given the information vector $\mathcal{I}_{i,\mathcal{N}_i}(t-1,t-1,t-1)$. 

Define $\hat{\bm{r}}_{y,i}(t)  \doteq  
\begin{bmatrix} 
\hat{\bm{x}}'_i(t) & \hat{\bm{x}}'_{\mathcal{N}_i}(t-1)  & \bm{y}'_i(t) 
\end{bmatrix}^{'}$. 
We make an approximation that given $\hat{\bm{r}}_{y,i}(t-1)$, $\bm{y}_i(t)$ is independent of the past information.
Let us write the conditional covariance of $\bm{y}_i(t)$ given the information at sensor $i$ as:
\begin{IEEEeqnarray}{rCl}\label{eqn:cond_cov_obs}
    \hat{\bm{P}}_{y,i}(t) &=& cov[\bm{y}_{i}(t) | \mathcal{I}_{i,\mathcal{N}_i}(t-1,t-1,t-1)] \nonumber\\
    &\approx& cov[\bm{y}_{i}(t)|\hat{\bm{x}}_{i}(t-1),\hat{\bm{x}}_{\mathcal{N}_i}(t-2),\bm{y}_{i}(t-1)] \nonumber\\
    &=& cov[\bm{y}_{i}(t)|\hat{\bm{r}}_{y,i}(t-1)]
\end{IEEEeqnarray} 

\noindent As argued in section \ref{subsection:conditional-covariance-of-estimate},
\begin{IEEEeqnarray}{rCl}
    &&cov[\bm{y}_{i}(t)|\hat{\bm{r}}_{y,i}(t-1)] \nonumber\\
    &=& cov[\bm{y}_{i}(t)] \nonumber\\
    &-&\scalemath{0.87}{cov[\bm{y}_{i}(t),\hat{\bm{r}}_{y,i}(t-1)] cov[\hat{\bm{r}}_{y,i}(t-1)]^{-1}cov[\hat{\bm{r}}_{y,i}(t-1),\bm{y}_{i}(t)]} \IEEEeqnarraynumspace
\end{IEEEeqnarray}

\noindent Now we find each term in the above expression using similar methods used in section \ref{subsection:conditional-covariance-of-estimate}.
Firstly, we have,
\begin{IEEEeqnarray*}{rCl}
    cov[\bm{y}_{i}(t)] &=& cov[\bm{A}\bm{x}(t) + \bm{w}(t)] \\
    &=& \bm{A}\bm{B}(t)\bm{A}' + \bm{Q}
\end{IEEEeqnarray*}

\noindent Next, we have 
\begin{IEEEeqnarray*}{rCl}
     &&cov[\bm{y}_{i}(t),\hat{\bm{r}}_{y,i}(t-1)]\nonumber\\ &=&\mathbb{E}\left(\bm{y}_{i}(t) 
    \begin{bmatrix} 
    \hat{\bm{x}}'_i(t-1) & \hat{\bm{x}}'_{\mathcal{N}_i}(t-2)  & \bm{y}'_i(t-1) 
    \end{bmatrix} 
    \right)\nonumber\\
    &=&
    \begin{bmatrix} 
    cov(\bm{y}_{i}(t), \hat{\bm{x}}_i(t-1)) \\
    cov(\bm{y}_{i}(t), \hat{\bm{x}}_{\mathcal{N}_i}(t-2)) \\
    cov(\bm{y}_{i}(t),\bm{y}_i(t-1)) 
    \end{bmatrix}^{'}\\
\end{IEEEeqnarray*}

\noindent Using $\bm{y}_i(t) = \bm{C}_i\bm{A} \bm{x}(t-1) + \bm{C}_i\bm{w}(t-1) + \bm{v}_i(t)$, we write,
\begin{IEEEeqnarray*}{rCl}
    &&cov(\bm{y}_{i}(t), \hat{\bm{x}}_i(t-1)) \\ 
    &=& cov(\bm{C}_i\bm{A} \bm{x}(t-1) + \bm{C}_i\bm{w}(t-1) + \bm{v}_i(t), \hat{\bm{x}}_i(t-1)) \\
    &=&\bm{C}_i\bm{A} \bm{H}'_i(t-1)\bm{A}'\bm{C}'_i 
\end{IEEEeqnarray*}

\noindent Next we find, for all $j \in \mathcal{N}_i$,  
\begin{IEEEeqnarray*}{rCl}
    &&cov(\bm{y}_{i}(t), \hat{\bm{x}}_j(t-2)) \\ 
    &=& \scalemath{0.82}{cov(\bm{C}_i\bm{A}^2 \bm{x}(t-2) + \bm{C}_i\bm{A}\bm{w}(t-2) + \bm{C}_i\bm{w}(t-1)+ \bm{v}_i(t), \hat{\bm{x}}_j(t-2))} \\
    &=&\bm{C}_i\bm{A}^2 \bm{H}'_j(t-2)
\end{IEEEeqnarray*}

\noindent Finally,
\begin{IEEEeqnarray*}{rCl}
    &&cov(\bm{y}_{i}(t), \bm{y}_{i}(t-1)) \\ 
    &=&\scalemath{0.85}{ cov(\bm{C}_i\bm{A} \bm{x}(t-1) + \bm{C}_i\bm{w}(t-1) + \bm{v}_i(t), \bm{C}_i\bm{x}(t-1) + \bm{v}_i(t-1))} \\
    &=&\bm{C}_i\bm{A} \bm{B}(t-1)\bm{C}'_i
\end{IEEEeqnarray*}

\noindent Now,
\begin{IEEEeqnarray}{rCl}
    &&cov[\hat{\bm{r}}_{y,i}(t-1) \nonumber\\
    &=&\mathbb{E}\left(\begin{pmatrix} 
    \hat{\bm{x}}_i(t-1) \\ \hat{\bm{x}}_{\mathcal{N}_i}(t-2) \\ \bm{y}_i(t-1)  
    \end{pmatrix} 
    \begin{pmatrix} \hat{\bm{x}}_i(t-1) \\ \hat{\bm{x}}_{\mathcal{N}_i}(t-2) \\ \bm{y}_i(t-1)  
    \end{pmatrix}^{'} \right) \nonumber\\
    &=&\scalemath{0.67}{\begin{bmatrix}
    \bm{L}_i(t-1) & \bm{J}_{i,{\mathcal{N}_i}}(t-1)&(\bm{H}_i(t-1)+ \bm{M}_i(t-1))\bm{C}'_i \\   \bm{J}'_{i,{\mathcal{N}_i}}(t-1) & \bm{T}_{\mathcal{N}_i,\mathcal{N}_i}(t-2) & \bm{H}_{\mathcal{N}_i}(t-2)\bm{A}'\bm{C}'_i \\ \bm{C}_i (\bm{H}'_i(t-1) +\bm{M}'_i(t-1))  & \bm{C}_i \bm{A} \bm{H}'_{\mathcal{N}_i}(t-2) & \bm{C}_i \bm{B}(t-1) \bm{C}'_i + \bm{R}_i
    \end{bmatrix}}
    \IEEEeqnarraynumspace
\end{IEEEeqnarray}

\noindent where, $\bm{J}_{i,\mathcal{N}_i}(t-1) = \begin{bmatrix} 
    \bm{J}'_{i,j_1}(t-1) \\ \vdots \\ \bm{J}'_{i,j_{N_i}}(t-1) 
    \end{bmatrix}^{'}$
    
\noindent and $\bm{J}_{i,j_1}(t-1) = cov( \hat{\bm{x}}_i(t-1),\hat{\bm{x}}_{j_1}(t-2))$ and can be calculated as,
\begin{IEEEeqnarray*}{rCl}
    && cov(\hat{\bm{x}}_i(t-1),\hat{\bm{x}}_{j_1}(t-2)) \\
    &=& \bm{G}_{i,t-1}\bm{H}'_{j_1}(t-2) + \bm{D}_{i,t-1}\bm{T}_{i,j_1}(t-2) \\
    &+& \gamma \bm{F}_{i,t-1}\sum_{r\in \mathcal{N}_i}T_{r,j_1}(t-2)
\end{IEEEeqnarray*}

Using the preceding equations, we can calculate the conditional covariance defined in \eqref{eqn:cond_cov_obs}.
The conditional mean of $\bm{y}_i(t)$ given $\hat{\bm{r}}_{y,i}(t-1)$ would be written similarly as in section \ref{subsection:conditional-expectation-of-estimate},
\begin{IEEEeqnarray*}{rCl}
    \bm{\hat{\mu}}_{y,i}(t) &\triangleq& \mathbb{E}[\bm{y}_{i}(t) | \mathcal{I}_{i,\mathcal{N}_i}(t-1,t-1,t-1)] \\
    &\approxeq& \mathbb{E}[\bm{y}_{i}(t)|\hat{\bm{r}}_{y,i}(t-1)]\\ &=&\scalemath{0.95}{cov(\hat{\bm{y}}_i(t),\hat{\bm{r}}_{y,i}(t-1))cov(\hat{\bm{r}}_{y,i}(t-1))^{-1} \hat{\bm{r}}_{y,i}(t-1)}
\end{IEEEeqnarray*}

\subsection{Pre and post change conditional distribution of $\hat{\bm{x}}_i(t)$}\label{subsection:Conditional-distribution-of-estimate}
\subsubsection{No attack}\label{subsubsection:Conditional-distribution-of-no-attack-estimate}
The \textbf{conditional} distribution of $\bm{\hat{x}}_i(t)$ given the information $\mathcal{I}_{i,\mathcal{N}_i}(t-1,t,t)$ in case of no attack is:
\begin{eqnarray}
     f_{\infty,i} \equiv  \mathcal{N}(\bm{\hat{\mu}}_i(t), \bm{\hat{P}}_i(t)) \label{eqn:conditional_distribution_of_unattacked_estimate}
\end{eqnarray}
where $f_{\infty}$ denotes the distribution under  $\tau = \infty$.

Also note that the conditional distribution of a neighbor's estimate at the current sensor can be denoted as,
\begin{align} \label{eqn:pre-att-cond-dist-neighbor-est}
p(\hat{\bm{x}}_j(t-1)|\mathcal{I}_{i,\mathcal{N}_i}(t-1,t-1,t)) \sim \mathcal{N}(\bm{\hat{\mu}}_j^i(t), \bm{\hat{P}}_j^i(t)) 
\end{align}

\subsubsection{Attack}\label{subsubsection:Conditional-distribution-of-Attack-estimate}
The error propagates across the network through information exchange between the neighbors. If the attack vector $\bm{e}_j(t)$ is injected at a sensor $j$, the covariance in \eqref{eqn:approximate-conditional-covariance-of-estimate} changes for all $i \in \mathcal{N}$. 
The exchange of information triggers error propagation to other nodes.

Let $\tilde{\bm{x}}_i(t)$ be the estimate after attack.
If the attack occurs at a unknown time, $\tau$, the conditional distribution of the estimate at sensor $i$ at time $t$ ($ \geq \tau$) given that the attack occurs at sensor $j$  can be written as:
\begin{eqnarray}
    f_{\tau,t,i,j}\equiv \mathcal{N}(\tilde{\bm{\mu}}_{i,j,\tau}(t), \tilde{\bm{P}}_{i,j,\tau}(t)) \label{eqn:conditional_distribution_of_attacked_estimate}
\end{eqnarray}

where $\tilde{\bm{\mu}}_{i,j,\tau}(t)$ and $\tilde{\bm{P}}_{i,j,\tau}(t))$ are the conditional mean and covariance of $\tilde{\bm{x}}_i(t)$, given sensor $j$ is under attack from time $\tau$ onward. The mean and the covariance change at each sensor after an attack is launched, because of message exchange in the KCIF. However, the sensor directly under attack will have a different change in the moments than other sensors where the attack only propagates due to information exchange. The observation noise covariance at the attacked sensor becomes $\bm{R}_j + \bm{\Sigma}$ which can be  incorporated in the conditional covariance and mean calculations in Section~\ref{subsection:conditional-covariance-of-estimate} and \ref{subsection:conditional-expectation-of-estimate} to find the post-attack distributions.

With the same arguments, we can write the post attack conditional distribution of the neighbor's estimate at sensor $i$ given attack occurs at sensor $j_1$ as,
\begin{align} \label{eqn:post-att-cond-dist-neighbor-est}
p(\tilde{\bm{x}}_j(t-1)|\mathcal{I}_{i,\mathcal{N}_i}(t-1,t-1,t),j_1) \sim \mathcal{N}(\tilde{\bm{\mu}}_{j,j_1}^i(t), \tilde{\bm{P}}_{j,j_1}^i(t)) 
\end{align}

Also note that the pre-attack conditional distribution of the observation at sensor $i$ can be written down as,
\begin{align}\label{eqn:pre-att-conditional_distribution_of_observation}
p(\bm{y}_{i}(t) | \mathcal{I}_{i,\mathcal{N}_i}(t-1,t-1,t-1)) \equiv  \mathcal{N}(\bm{\hat{\mu}}_{y,i}(t), \bm{\hat{P}}_{y,i}(t)) 
\end{align}

The post attack conditional distribution of the observation at sensor $i$ given attack occurs at sensor $j$ can be given as,
\begin{align}\label{eqn:post-att-conditional_distribution_of_observation}
p(\tilde{\bm{y}}_{i}(t) | \mathcal{I}_{i,\mathcal{N}_i}(t-1,t-1,t-1),j) \equiv  \mathcal{N}(\tilde{\bm{\mu}}_{y,i,j}(t), \tilde{\bm{P}}_{y,i,j}(t)) 
\end{align}

\begin{figure*}
\hrule
\footnotesize
\begin{IEEEeqnarray}{rCL} \label{eqn:recursive lambda}
    && \lambda_i^l(t) \nonumber \\
    &=& \frac{\mathbb{P}(\tau\leq t)}{\mathbb{P}(\tau>t)} \frac{\left( \frac{\mathbb{P}(\tau\leq t-1)}{\mathbb{P}(\tau\leq t)}p(\mathcal{I}_{i,\mathcal{N}_i}(t-1,t-1,t-1)|\tau\leq t-1,l) + \frac{\mathbb{P}(\tau=t)}{\mathbb{P}(\tau\leq t)}p(\mathcal{I}_{i,\mathcal{N}_i}(t-1,t-1,t-1)|\tau=t,l)\right)}{p(\mathcal{I}_{i,\mathcal{N}_i}(t-1,t-1,t-1)|\tau > t-1,l)} \times \nonumber\\ &&\frac{p(\bm{\hat{x}}_i(t)|\mathcal{I}_{i,\mathcal{N}_i}(t-1,t,t),\tau\leq t,l) \displaystyle \prod_{j \in \mathcal{N}_i}p(\bm{\hat{x}}_j(t-1)|\mathcal{I}_{i,\mathcal{N}_i}(t-1,t-1,t),\tau\leq t,l)p(\bm{y}_i(t) | \mathcal{I}_{i,\mathcal{N}_i}(t-1,t-1,t-1), \tau \leq t,l)}{ p(\bm{\hat{x}}_i(t)|\mathcal{I}_{i,\mathcal{N}_i}(t-1,t,t),\tau> t) \displaystyle \prod_{j \in \mathcal{N}_i}p(\bm{\hat{x}}_j(t-1)|\mathcal{I}_{i,\mathcal{N}_i}(t-1,t-1,t),\tau>t)p(\bm{y}_i(t) | \mathcal{I}_{i,\mathcal{N}_i}(t-1,t-1,t-1), \tau > t)} \nonumber\\
    &=& 
    \left(\frac{
    \mathbb{P}(\tau\leq t-1)p(\mathcal{I}_{i,\mathcal{N}_i}(t-1,t-1,t-1)|\tau\leq t-1,l)}
    {(1-\rho)\mathbb{P}(\tau>t-1)p(\mathcal{I}_{i,\mathcal{N}_i}(t-1,t-1,t-1)|\tau> t-1,l)} + \frac{\mathbb{P}(\tau=t)}{(1-\rho)\mathbb{P}(\tau>t-1)}\right) \times \nonumber\\ 
    &&\frac {p(\bm{\hat{x}}_i(t)|\mathcal{I}_{i,\mathcal{N}_i}(t-1,t,t),\tau\leq t,l) \displaystyle \prod_{j \in \mathcal{N}_i}p(\bm{\hat{x}}_j(t-1)|\mathcal{I}_{i,\mathcal{N}_i}(t-1,t-1,t),\tau\leq t,l)p(\bm{y}_i(t) | \mathcal{I}_{i,\mathcal{N}_i}(t-1,t-1,t-1), \tau \leq t,l)}{p(\bm{\hat{x}}_i(t)|\mathcal{I}_{i,\mathcal{N}_i}(t-1,t,t),\tau> t) \displaystyle \prod_{j \in \mathcal{N}_i}p(\bm{\hat{x}}_j(t-1)|\mathcal{I}_{i,\mathcal{N}_i}(t-1,t-1,t),\tau>t)p(\bm{y}_i(t) | \mathcal{I}_{i,\mathcal{N}_i}(t-1,t-1,t-1), \tau > t)}  \nonumber\\
    &=& \scalemath{0.95}{\left(\frac{\lambda_i^l(t-1)+\rho}{1-\rho} \right)\frac{p(\bm{\hat{x}}_i(t)|\mathcal{I}_{i,\mathcal{N}_i}(t-1,t,t),\tau\leq t,l) \displaystyle \prod_{j \in \mathcal{N}_i}p(\bm{\hat{x}}_j(t-1)|\mathcal{I}_{i,\mathcal{N}_i}(t-1,t-1,t),\tau\leq t,l)p(\bm{y}_i(t) | \mathcal{I}_{i,\mathcal{N}_i}(t-1,t-1,t-1), \tau \leq t,l)}{p(\bm{\hat{x}}_i(t)|\mathcal{I}_{i,\mathcal{N}_i}(t-1,t,t),\tau> t) \displaystyle \prod_{j \in \mathcal{N}_i}p(\bm{\hat{x}}_j(t-1)|\mathcal{I}_{i,\mathcal{N}_i}(t-1,t-1,t),\tau>t)p(\bm{y}_i(t) | \mathcal{I}_{i,\mathcal{N}_i}(t-1,t-1,t-1), \tau > t)}} \IEEEeqnarraynumspace
\end{IEEEeqnarray}
\normalsize
\hrule
\end{figure*}

\section{The Bayesian Quickest Detection}\label{section:Bayesian Detection}
In this section, we address the problem of Bayesian quickest detection of FDI to identify an attack starting at an unknown time $\tau$, which follows a geometric distribution with a mean of $\frac{1}{\rho}$, a known parameter known to all the sensors. Our objective is to find an optimal stopping time $T$ that minimizes the expected detection delay while maintaining a constraint on the probability of false alarm. It is important to note that, in our model, each sensor detects the change. The Bayesian quickest detection  problem involves a test similar to Shriyaev's test for quickest detection \cite{poor2008quickest} at each sensor. We formulate the problem at any sensor  as follows:
\begin{align}
\min_T {\mathbb{E}[(T-\tau)^+]} \quad \textbf{subject to} \quad \mathbb{P}(T<\tau) \leq \alpha
\end{align}
where $\mathbb{E}[(T-\tau)^+]$ represents the expected detection delay, and $\mathbb{P}(T<\tau)$ denotes the probability of false alarm, constrained to be less than or equal to $\alpha$.

The optimal stopping time (and also rule with a a little abuse of notation) at sensor $i$, denoted by $T_i$, is determined based on the posterior distribution of the attack instant, $\pi_i(t)$. Here, $\pi_i(t)$ represents the probability that the attack has occurred by time $t$, given all the available information at sensor $i$ up to time $t$. Each sensor independently makes its decision on attack detection using its local information.

To test for the hypotheses $H_0: \tau > t$ versus $H_1: \tau \leq t$ at time $t$ at sensor $i$, we use the following expression for the optimal stopping time:
\begin{align}\label{eqn:optimal-stopping-time}
T_i = \inf \{t\geq1|\pi_i(t) \geq \Lambda_i\}
\end{align}
where, $\Lambda_i>0$ is a local threshold parameter.

The computation of $\pi_i(t)$ relies on the posterior probability of the attack instant, which is expressed as follows:
\begin{align}\label{eqn:posterior-of-change-instant}
\pi_i(t) &\doteq \max_{l \in \mathcal{N}}\mathbb{P}(\tau \leq t|\mathcal{I}_{i,\mathcal{N}_i}(t,t,t),l) \nonumber\\
&= \max_{l \in \mathcal{N}} \pi_i^l(t)
\end{align}

Here $\mathcal{I}_{i,\mathcal{N}_i}(t,t,t)$ represents the information available to sensor $i$ at time $t$ and $l$ is the hypothesized sensor under attack. $\pi_i^l(t)$ is the posterior of the change point at sensor $i$ given attack occurs at sensor $l$. Obviously, the maximizing $l$ in \eqref{eqn:posterior-of-change-instant} will be treated as the attacked node by the $i$-th sensor, if $\pi_i(t) \geq \Lambda_i$.

It is important to note that, unlike in general Bayesian QCD problems, the  challenge lies in dealing with non-i.i.d observations pre and post attack, making the derivation of a recursive relation for $\pi_i^l(t)$ non-trivial.
\subsection{Computation of $\pi_i^l(t)$}\label{subsection:Computation of pi_i^l(t)}
Note that the attack instant is independent of the sensor under attack. We can then rewrite \eqref{eqn:posterior-of-change-instant} as:
\begin{eqnarray}\label{eqn:pi_lambda_relation}
    \pi_i^l(t) =& \frac{\mathbb{P}(\tau\leq t)p(\mathcal{I}_{i,\mathcal{N}_i}(t,t,t)|\tau\leq t,l)}{\mathbb{P}(\tau\leq t)p(\mathcal{I}_{i,\mathcal{N}_i}(t,t,t)|t\leq \tau,l) + \mathbb{P}(\tau> t)p(\mathcal{I}_{i,\mathcal{N}_i}(t,t,t)|\tau>t,l)} \nonumber\\
    =&  \frac{\lambda_i^l(t)}{1+\lambda_i^l(t)}
\end{eqnarray}
where
\begin{align}\label{eqn:lambda} 
   \lambda_i^l(t) \doteq \frac{\mathbb{P}(\tau\leq t)p(\mathcal{I}_{i,\mathcal{N}_i}(t,t,t)|\tau\leq t,l)}{\mathbb{P}(\tau>t)p(\mathcal{I}_{i,\mathcal{N}_i}(t,t,t)|\tau>t,l)} 
\end{align}

We look at the denominator first in \eqref{eqn:lambda}:
\begin{eqnarray}\label{eqn:pre_attack_distribution_of_information}
     && p(\mathcal{I}_{i,\mathcal{N}_i}(t,t,t)|\tau>t,l) \nonumber\\
     &=& p(\bm{\hat{x}}_i(t),\mathcal{I}_{i,\mathcal{N}_i}(t-1,t,t)|\tau>t,l) \nonumber\\
     &=& \scalemath{0.95}{p(\bm{\hat{x}}_i(t)|\mathcal{I}_{i,\mathcal{N}_i}(t-1,t,t),\tau>t) p(\mathcal{I}_{i,\mathcal{N}_i}(t-1,t,t)|\tau>t,l)} \nonumber
\end{eqnarray}

The first term in \eqref{eqn:pre_attack_distribution_of_information} is the pre-attack distribution, thus does not depend on the attacked sensor and can be found using \eqref{eqn:conditional_distribution_of_unattacked_estimate}. The last term can be written as:
\begin{eqnarray}\label{eqn:prob_preattackinfo}
     && p(\mathcal{I}_{i,\mathcal{N}_i}(t-1,t,t)|\tau>t,l) \nonumber\\
     &=& p(\bm{\hat{x}}_{\mathcal{N}_i}(t-1),\mathcal{I}_{i,\mathcal{N}_i}(t-1,t-1,t)|\tau>t,l) \nonumber\\
     &=& p(\bm{\hat{x}}_{\mathcal{N}_i}(t-1)|\mathcal{I}_{i,\mathcal{N}_i}(t-1,t-1,t),\tau>t,l) \times \nonumber\\ &&p(\mathcal{I}_{i,\mathcal{N}_i}(t-1,t-1,t)|\tau>t,l) \nonumber\\
     &\approxeq& \prod_{j \in \mathcal{N}_i }(p(\bm{\hat{x}}_j(t-1)|\mathcal{I}_{i,\mathcal{N}_i}(t-1,t-1,t),\tau>t)) \times \nonumber\\
     && p(\mathcal{I}_{i,\mathcal{N}_i}(t-1,t-1,t)|\tau>t,l)
\end{eqnarray}
where at sensor $i$, we make an approximation that the estimates of neighbor's of sensor $i$ are conditionally independent and therefore can write the product form as shown above. 

 The first term in \eqref{eqn:prob_preattackinfo} can be found using \eqref{eqn:pre-att-cond-dist-neighbor-est} by ignoring the neighbors of sensor $j \in \mathcal{N}_i$ other than $i$ while deriving the conditional distribution of $\bm{\hat{x}}_j(\cdot)$ at sensor $i$ given the information available at sensor $i$. Also note that,  before the commencement of attack, this distribution is independent of the attacked sensor. We also assume that all nodes know the parameters of their neighboring nodes.

The last term in \eqref{eqn:prob_preattackinfo} can be further simplified as:
\begin{eqnarray}\label{eqn:temporary_dist_observation}
    && p(\mathcal{I}_{i,\mathcal{N}_i}(t-1,t-1,t)|\tau > t,l) \nonumber\\
    &=& p(\bm{y}_i(t),\mathcal{I}_{i,\mathcal{N}_i}(t-1,t-1,t-1)|\tau > t,l)   \nonumber\\
    &=& p(\bm{y}_i(t) | \mathcal{I}_{i,\mathcal{N}_i}(t-1,t-1,t-1), \tau > t) \times \nonumber\\
    && p(\mathcal{I}_{i,\mathcal{N}_i}(t-1,t-1,t-1)|\tau > t-1,l)
\end{eqnarray}

The first term in \eqref{eqn:temporary_dist_observation} can be found using \eqref{eqn:pre-att-conditional_distribution_of_observation}. Also pre-attack, we drop the attacked sensor index from the distribution of $\bm{y}_i(t)$. The second term can be calculated recursively from  \eqref{eqn:pre_attack_distribution_of_information}.

Now we consider the numerator in \eqref{eqn:lambda}:
\begin{IEEEeqnarray}{rCl}\label{eqn:post_attack_distribution_of_information}
    && p(\mathcal{I}_{i,\mathcal{N}_i}(t,t,t)|\tau \leq t,l) \nonumber\\
    &=& p(\bm{\hat{x}}_i(t),\mathcal{I}_{i,\mathcal{N}_i}(t-1,t,t)|\tau \leq t,l)   \nonumber\\
     &=& \scalemath{0.91}{p(\bm{\hat{x}}_i(t)|\mathcal{I}_{i,\mathcal{N}_i}(t-1,t,t),\tau\leq t,l) p(\mathcal{I}_{i,\mathcal{N}_i}(t-1,t,t)|\tau \leq t,l)} \IEEEeqnarraynumspace
\end{IEEEeqnarray}

\noindent where,
\begin{eqnarray}
    && p(\mathcal{I}_{i,\mathcal{N}_i}(t-1,t,t)|\tau \leq t,l) \nonumber\\
    &=& p(\bm{\hat{x}}_{\mathcal{N}_i}(t-1),\mathcal{I}_{i,\mathcal{N}_i}(t-1,t-1,t)|\tau \leq t,l) \nonumber\\
    &\approxeq& \prod_{j \in \mathcal{N}_i }(p(\bm{\hat{x}}_j(t-1)|\mathcal{I}_{i,\mathcal{N}_i}(t-1,t-1,t),\tau \leq t,l)) \times \nonumber\\
    && p(\mathcal{I}_{i,\mathcal{N}_i}(t-1,t-1,t)|\tau \leq t,l)
\end{eqnarray}

The above approximation is similar to the one used in \eqref{eqn:prob_preattackinfo}. The last term in the above equation can be further simplified as
\begin{eqnarray}\label{eqn:temp_dist_obs_att}
    && p(\mathcal{I}_{i,\mathcal{N}_i}(t-1,t-1,t)|\tau \leq t,l) \nonumber\\
    &=& p(\bm{y}_i(t),\mathcal{I}_{i,\mathcal{N}_i}(t-1,t-1,t-1)|\tau \leq t,l)   \nonumber\\
    &=& p(\bm{y}_i(t) | \mathcal{I}_{i,\mathcal{N}_i}(t-1,t-1,t-1), \tau \leq t,l) \times \nonumber\\
    && p(\mathcal{I}_{i,\mathcal{N}_i}(t-1,t-1,t-1)|\tau \leq t,l) 
\end{eqnarray}

The first term in \eqref{eqn:temp_dist_obs_att} depends on the attacked sensor and can be found using \eqref{eqn:post-att-conditional_distribution_of_observation}.

Also, 
\begin{IEEEeqnarray}{rCl}
    && p(\mathcal{I}_{i,\mathcal{N}_i}(t-1,t-1,t-1)|\tau \leq t,l) \nonumber\\
    &=& \scalemath{0.95}{p(\mathcal{I}_{i,\mathcal{N}_i}(t-1,t-1,t-1)|\tau\leq t-1,l) \mathbb{P}(\tau\leq t-1|\tau\leq t)} \nonumber\\
    &+& p(\mathcal{I}_{i,\mathcal{N}_i}(t-1,t-1,t-1)|\tau=t,l)\mathbb{P}(\tau=t|\tau\leq t)
\end{IEEEeqnarray}

 The term $p(\bm{\hat{x}}_i(t)|\mathcal{I}_{i,\mathcal{N}_i}(t-1,t,t),\tau\leq t,l)$ in \eqref{eqn:post_attack_distribution_of_information} is the post change conditional distribution of the estimate at sensor $i$ and can be written down as
\begin{IEEEeqnarray}{rCl} \label{eqn:kth_term_in_recursion}
    && p(\bm{\hat{x}}_i(t)|\mathcal{I}_{i,\mathcal{N}_i}(t-1,t,t),\tau\leq t,l) \nonumber\\
    &=&\scalemath{0.95}{\sum_{m=0}^{t}p(\bm{\hat{x}}_i(t)|\mathcal{I}_{i,\mathcal{N}_i}(t-1,t,t),\tau=m,l)\mathbb{P}(\tau=m|\tau\leq t)} \IEEEeqnarraynumspace
\end{IEEEeqnarray}

where the distribution $p(\bm{\hat{x}}_i(t)|\mathcal{I}_{i,\mathcal{N}_i}(t-1,t,t),\tau=m,l)$ can be found in Section~\ref{subsubsection:Conditional-distribution-of-Attack-estimate} with $l$ as the attacked sensor.

\noindent Using \eqref{eqn:pre_attack_distribution_of_information}, \eqref{eqn:post_attack_distribution_of_information} in \eqref{eqn:lambda}, a recursive relation for $\lambda_i^l(t)$ can be obtained as seen in \eqref{eqn:recursive lambda}. 

\section{Non-Bayesian QCD}\label{section:Non-Bayesian QCD}
In this section, we solve the distributed non-Bayesian quickest detection-isolation problem for known and unknown $\bm{\Sigma}$. 
 The attack is initiated at  an arbitrary unknown  instant $\tau$. In the non-i.i.d. observation case, Lai \cite{lai1998information} proposed a generalized CUSUM test for quickest attack detection, showing it's asymptotic optimality. However we view our QCD problem at any node as a multiple hypothesis testing problem where we can have multiple post change hypotheses, each representing one possible attacked sensor. We seek to detect the change due to attack as quickly as possible and also identify the sensor under attack. In the literature the problem is referred to as joint change point detection-identification problem, and has been studied in \cite{nikiforov2002optimal,tartakovsky2008multidecision, lai2000sequential}. Motivated by these works, we seek to minimize the worst case mean delay for detection-isolation:
\begin{align*}
    \bar{\mathbb{E}}(T_i) = \max_{j \in \mathcal{N}}\sup_{\tau\geq1}\esssup(\mathbb{E}_{\tau}^{j}(T_i-\tau|T_i\geq \tau),
\end{align*}

where the conditional expectation given the attack having been started at time $\tau$ at node $j$ is denoted by $\mathbb{E}_{\tau}^{j}$,  $\mathbb{P}_{\tau}^{j}$ denotes the corresponding conditional probability measure, and $T_i$ is the stopping time for detection by sensor $i$.

The $i$-th node seeks to minimize the worst case average detection delay subject to  constraints on the average run length to false alarm (ARL2FA) and the probability of misidentification:
\begin{align} \label{eqn:lai's problem}
    &\inf_{\delta_i}\bar{\mathbb{E}}(T_i) \nonumber\\
    &\text{s.t.} \quad \mathbb{E}_{\infty}[T_i]\geq \gamma \quad \text{and} \quad \max_{j\neq i}\sup_{\tau \geq 1}\mathbb{P}_{\tau}^{j}(d_i\neq j|T_i\geq \tau) \leq \beta
\end{align}

where $\delta_i=(T_i,d_i)$ is the detection-identification rule at sensor $i$, and $d_i$ is the decision (rule) to identify the sensor under attack. On the other hand,  $\mathbb{E}_{\infty}(\cdot)$ denotes the expectation under no change, i.e., given that $\tau=\infty$.

\subsection{Known $\bm{\Sigma}$}
It was shown by Lai \cite{lai2000sequential} that under certain assumptions, the windowed multiple hypotheses sequential probability ratio test (MSPRT) turns out to be asymptotically optimal for the problem defined in \eqref{eqn:lai's problem}. In order to adapt the windowed MSPRT to our problem, the log likelihood ratio at node $i$ at time $t$ is defined as: 
\begin{align*}
    L_{i,j}^{t,\tau} = log\left(\frac{f_{\tau,t,i,j}(\bm{\hat{x}}_i(t)|\mathcal{I}_{i,\mathcal{N}_i}(t-1,t,t))}{f_{\infty,i}(\bm{\hat{x}}_i(t)|\mathcal{I}_{i,\mathcal{N}_i}(t-1,t,t))}\right)
\end{align*}

The probability distributions can be calculated using \eqref{eqn:conditional_distribution_of_unattacked_estimate} and \eqref{eqn:conditional_distribution_of_attacked_estimate} of section \ref{section:Dynamics of estimate}. 
The windowed-MSPRT algorithm at sensor $i$ is then defined as:
\begin{align*}
    T_i = \inf\{n\geq1: \max_{n-t_{\gamma} \leq k \leq n} \min_{1\leq j \leq N, j\neq i}\sum_{t=k}^{n} L_{i,j}^{t,k} \geq b_i\},
\end{align*}

where $T_i$ is the MSPRT stopping time at sensor $i$, and $b_i>0$ is a local positive threshold adapted to meet the ARL2FA constraint. The window size is denoted by $t_{\gamma}$ and  is chosen as  $t_{\gamma}=\mathcal{O}(\log{\gamma})$.


Unfortunately, the above statistic is non-recursive\cite{nikiforov2002optimal} and with increasing number of sensors, the computational complexity increases.

\subsection{Unknown $\bm{\Sigma}$}
Due to   unknown $\bm{\Sigma}$, which is assumed to belong to a known set $\bm{\Theta}$ of positive semi-definite matrices, it is natural to use a generalised likelihood ratio (GLR) based detection technique. To this end, we   use   the WL-GLR algorithm from \cite[Section~III-B]{lai1998information} designed for QCD under non-i.i.d. observations with unknown post-change parameters. 
The stopping time in WL-GLR\cite{lai1998information} is defined as:
\begin{IEEEeqnarray*}{rCl}
    T_i &=& \inf \{t\geq1| S_{i}^{t} \geq b_i\}\\
    \text{where} \\
    S_{i}^{t} &=& \scalemath{0.75}{\max_{1\leq j\leq N,j\neq i} \max_{t-t_{\gamma}\leq k \leq t}\sup_{\bm{\Sigma} \in \bm{\Theta}} \sum_{t'=k}^{t} log\frac{f_{k,t',i,j}(\bm{\hat{x}}_i(t')|\mathcal{I}_{i,\mathcal{N}_i}(t'-1,t',t');\bm{\Sigma})}{f_{\infty,i}(\bm{\hat{x}}_i(t')|\mathcal{I}_{i,\mathcal{N}_i}(t'-1,t',t'))}} 
\end{IEEEeqnarray*}

The window size   $t_{\gamma}$ is chosen as $t_{\gamma}=\mathcal{O}(\log\gamma)$. Obviously, the maximizing $j$ is used in attacked node identification. However, this maximization operation prohibits the test statistic from being updated recursively, unlike the standard WL-GLR algorithm for non-i.i.d. observations used in \cite{lai1998information}.  However, it can be updated recursively for each $\bm{\Sigma} \in \bm{\Theta}$ if we skip the maximization step over $j$ (i.e., the identification task) \cite{lai1998information}.

\section{Numerical Results}\label{section:Numerical Results}

We consider a network comprising five sensors, arranged according to the topology depicted in Figure 1. The dimension of the process is $p=2$ and the observation taken at each sensor is also two-dimensional with $q=2$. We set the consensus parameter $\gamma= 0.05$. The entries of matrices $\bm{A}$, $\bm{Q}$, $\bm{C_i}$, and $\bm{R_i}$ are  generated uniformly at random from the interval $[0, 1]$, while ensuring positive definiteness of $\bm{Q}$ and $\bm{R_i}$. The attacked sensor is chosen to be sensor $2$ with $\bm{\Sigma} = 3\bm{I}$.

\begin{figure}[htbp]
\centering
\includegraphics[height=5cm, width=2.5in]{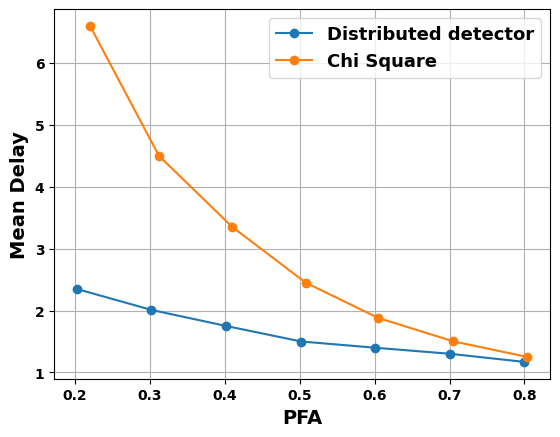}
\caption{Delay vs PFA for the Bayesian setting at sensor 1 with attack at sensor 2}
\label{fig:result_B}
\end{figure}

\begin{figure}[htbp]
\centering
\includegraphics[height=5cm, width=2.5in]{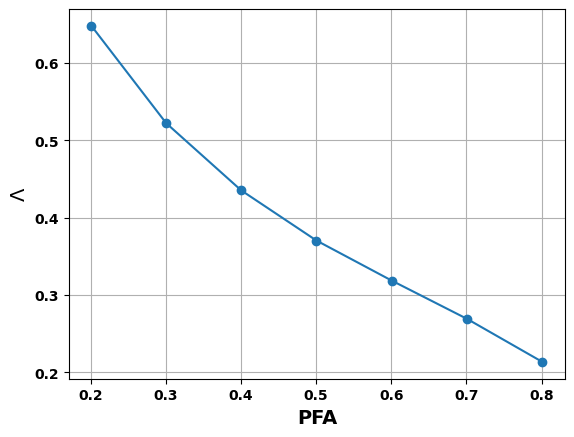}
\caption{Threshold vs PFA at the attacked sensor (sensor 2) for distributed detector in the Bayesian case}
\label{fig:lambdavspfa}
\end{figure}

For the Bayesian case, we assume $\rho=0.05$. To evaluate the performance of our distributed QCD algorithm, we first generate the process, observations, and attacks for $2500$ sample paths, each with a horizon length of $125$ time steps. We apply the Kalman consensus information filter at each sensor to calculate posterior probabilities using (\ref{eqn:recursive lambda}). 
For each sensor $i$, we dynamically update the detection threshold $\Lambda_i$ for every sample path using an online gradient descent algorithm:
\begin{align*}
    \Lambda_i(j) := \Lambda_i(j-1) + a(j)\times(\mathbbm{1}_{FA}^i(j) - \alpha),
\end{align*}
where $\Lambda_i(j)$ and $\mathbbm{1}_{FA}^i(j)$ represent the threshold and false alarm indicator for the $j^{th}$ sample path at sensor $i$, respectively. The sequence $a(j)$ adheres to the standard conditions of stochastic gradient descent algorithms: $\sum_{j=0}^{\infty}a(j)=\infty$ and $\sum_{j=0}^{\infty}a^{2}(j)<\infty$. The desired probability of false alarm is denoted by $\alpha$. This threshold is employed for attack detection, and unless a false alarm is triggered, the detection time delay is recorded. 
For comparison, we employ the windowed $\chi^2$ detector at each sensor, with the detection threshold determined through the same gradient descent process as used in our quickest detection algorithm. The detection window is set to $J=3$. The simulation outcomes demonstrate the superior performance of our distributed detector compared to the $\chi^2$ detector. Figure \ref{fig:result_B} shows that significant reduction in mean detection delay is possible for the same probability of false alarm (PFA) when one uses our distributed QCD algorithm instead of  the $\chi^2$ detector. It is important to note that while this figure specifically illustrates the results from Sensor 1, similar trends were observed across all sensors in our network. Figure \ref{fig:lambdavspfa} shows that the threshold decreases with PFA, which is intuitive.

\begin{figure}[H]
\centering
\includegraphics[height=5cm, width=2.5in]{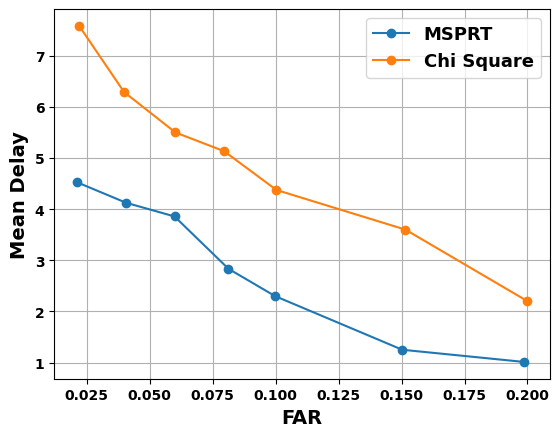}
\caption{Delay vs FAR for the Non-Bayesian case at sensor 1 with attack at sensor 2}
\label{fig:result_NB}
\end{figure}

\begin{figure}[H]
\centering
\includegraphics[height=5cm, width=2.5in]{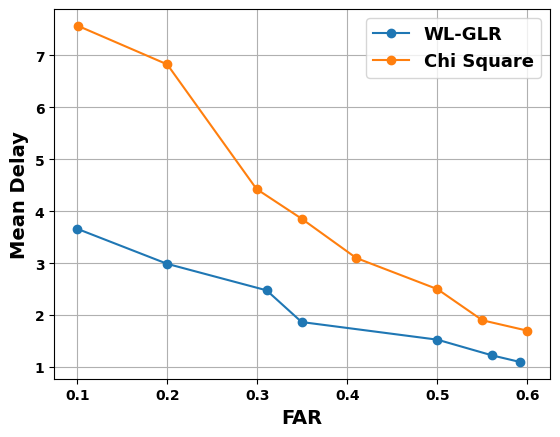}
\caption{Delay vs FAR for the Non-Bayesian case with unknown attack covariance at sensor 1 with attack at sensor 2}
\label{fig:result_NB_unknown}
\end{figure}

In the non-Bayesian case with known $\bm{\Sigma}$, Figure \ref{fig:result_NB} shows the variation of mean detection delay against false alarm rate (FAR). It is evident that the distributed MSPRT significantly outperforms the $\chi^2$ detector. Additionally, for unknown $\bm{\Sigma}$, Figure \ref{fig:result_NB_unknown} shows that WL-GLR significantly outperforms the $\chi^2$ detector.

\section{Conclusion}\label{section:Conclusion}

In this paper, we have presented a quickest detection algorithm for false data injection attack against a distributed tracking system employing the Kalman consensus information filter. Our algorithms for both Bayesian and non-Bayesian cases demonstrate superior performance over  $\chi^2$ detector. However, this algorithm can be extended in multiple directions such as  accounting for attacks across multiple sensors, handling unknown attack covariance in the Bayesian setting, and exploring the complexities of non-additive and nonlinear  attacks.

\small{
\bibliographystyle{IEEEtran}
\bibliography{refs}
}

\end{document}